# Reusing Static Analysis across Different Domain-Specific Languages using Reference Attribute Grammars

Johannes Mey[a], Thomas Kühn[b], René Schöne[a], and Uwe Aßmann[a]

a   Technische Universität Dresden, Germany
b   Karlsruhe Institute of Technology, Germany

**Abstract**

*Context:* Domain-specific languages (DSLs) enable domain experts to specify tasks and problems themselves, while enabling static analysis to elucidate issues in the modelled domain early. Although language workbenches have simplified the design of DSLs and extensions to general purpose languages, static analyses must still be implemented manually.

*Inquiry:* Moreover, static analyses, e.g., complexity metrics, dependency analysis, and declaration-use analysis, are usually domain-dependent and cannot be easily reused. Therefore, transferring existing static analyses to another DSL incurs a huge implementation overhead. However, this overhead is not always intrinsically necessary: in many cases, while the concepts of the DSL on which a static analysis is performed are *domain-specific*, the underlying algorithm employed in the analysis is actually *domain-independent* and thus can be reused in principle, depending on *how* it is specified. While current approaches either implement static analyses internally or with an external *Visitor*, the implementation is tied to the language's grammar and cannot be reused easily. Thus far, a commonly used approach that achieves reusable static analysis relies on the transformation into an intermediate representation upon which the analysis is performed. This, however, entails a considerable additional implementation effort.

*Approach:* To remedy this, it has been proposed to map the necessary domain-specific concepts to the algorithm's domain-independent data structures, yet without a practical implementation and the demonstration of reuse. Thus, we employ relational *Reference Attribute Grammars* (RAGs) by creating such a mapping to a domain-independent overlay structure using higher-order attributes.

*Knowledge:* We describe how static analysis can be specified on *analysis-specific* data structures, how relational RAGs can help with the specification, and how a mapping from the domain-specific language can be performed. Furthermore, we demonstrate how a static analysis for a DSL can be externalized and reused in another general purpose language.

*Grounding:* The approach was evaluated using the RAG system *JastAdd*. To illustrate reusability, we implemented two analyses with two addressed languages each: (1) a cycle detection analysis used in a small state machine DSL and for detecting circular dependencies between Java types and packages, as well as (2) an analysis of variable shadowing applied to both Java and the Modelica modelling language. Thereby, we demonstrate the reuse of two analysis algorithms in three completely different domains. Additionally, we use the cycle detection analysis to evaluate the efficiency by comparing our external analysis to an internal reference implementation analysing all Java programs in the *Qualitas Corpus*. Our evaluation indicates that an externalized analysis incurs only minimal overhead.

*Importance:* We make static analysis reusable for both DSLs and general purpose languages, showing the practicality and efficiency of externalizing static analysis using relational RAGs.

## The Art, Science, and Engineering of Programming

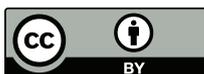

## 1  Introduction

Employing state-of-the-art language workbenches, the design of complex, custom *domain-specific languages* (DSLs) became feasible for researchers and practitioners alike [15]. As they encode domain-specific concepts and knowledge, DSLs enable domain experts to specify tasks and problems themselves. Moreover, DSLs allow for performing static analysis, e.g., complexity metrics, dependency analysis, and declaration-use analysis, on specified tasks and problems to discover issues in the modelled domain early. However, with the complexity of the DSL the effort for implementing static analysis increases, as well. Although the underlying algorithms of most static analysis are typically domain-independent, their implementation is tied to the DSL's concepts and relations. Thus, a static analysis implemented for one DSL cannot be easily reused for another DSL. Figure 1 depicts the four major approaches applied to reduce the effort for implementing and reusing static analysis.

Classic implementations employ an *External Visitor* [34], which supports adding new static analysis to a given DSL without changing the DSL's implementation. While external visitors reuse the traversal of the tree, these visitors depend on the domain-specific concepts of the DSL, which, in turn, prevents reusing visitor implementations.

An approach to reduce the implementation effort for visitors employs *Reference Attribute Grammars* (RAGs) to specify the static analysis by means of attributes and references (cf. section 2.3). These are then woven into only those types of concepts relevant for the analysis. In addition to simplifying the implementation, this approach enables incremental analysis, as changes to a program can be propagated while unchanged results are cached [40]. However, the implemented analysis is still woven into the domain-specific types and usually requires implementing custom traversals to resolve references. Thus, RAG-based static analyses are still hard to reuse.

By contrast, a practical approach to completely reuse static analyses between different DSLs is to introduce a common *Intermediate Representation* (IR) [29]. Static analysis can then be implemented only dependent on that IR. To reuse this analysis, a transformation from the DSL to the IR must be developed. Granted this approach permits completely reusing static analyses, yet, it requires the specification of a complete transformation, regardless of whether the specific analysis requires the complete transformation or not. Yet, implementing and maintaining an IR is a considerable effort, as it must provide suitable representations for all domain concepts of multiple DSLs. Thus, the achieved reusability is paid by a considerable implementation effort.

As an alternative, Joao Saraiva [36] proposed to employ higher-order attributes to achieve reusability of RAG-based static analysis. He proposed to use higher-order attributes to map the DSL's concepts to the analysis' concepts. While creating the latter, however, a reference to each domain-specific concept is kept by means of an identifier, which can be used to lookup the corresponding DSL's concept. The only downside to Saraiva's approach was the lack of a practical evaluation of the reusability and performance of the approach. In particular, while he pioneered the approach, he could not show its practical application and evaluate its performance overhead. Consequently, we illustrate the practical implementation of reusable static analyses for DSLs, whereas an analysis is implemented as a RAG specifically tailored

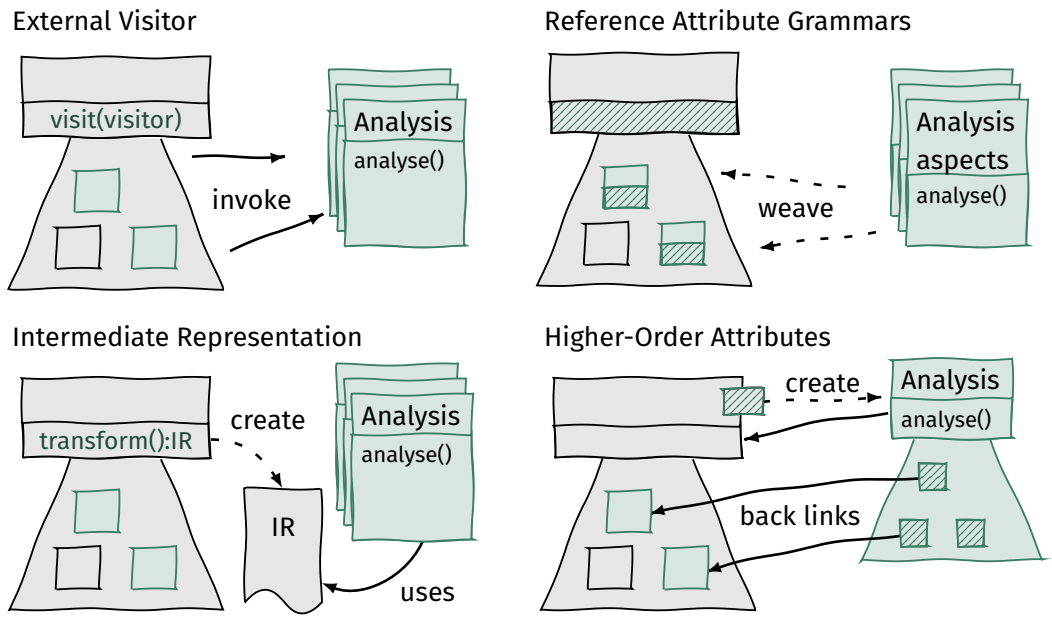

■ **Figure 1** Different approaches for implementing static analysis.

to the required information. To then reuse this analysis, only a higher-order attribute must be implemented mapping the relevant domain-specific concepts of the DSL to create the analysis' domain-independent concepts. Recent advances in RAGs, namely the introduction of *Relational Reference Attribute Grammars* in [32], enables us to overcome Saraiva limitation, by directly establishing and maintaining backlinks from domain-independent to domain-specific concepts. This reduces the effort of adding static analysis to DSLs and enables completely reusing existing domain-independent algorithms with limited performance overhead. We demonstrate our approach by developing and reusing two commonly used analyses, i.e., cycle detection and a variable shadowing analysis, for a *state machine* DSL and *Java*, as well as Modelica[4] and Java, respectively. This reusability is achieved by mapping the domain-specific concepts of these languages to a domain-independent RAG-based dependency graph and a definition-scope-tree. Moreover, to evaluate the incurred performance overhead, we employ the *Qualitas Corpus* [41] and compare our reusable cycle detection with a RAG-based implementation. In conclusion, this paper shows that reusable static analyses are both feasible and practical.

## 2    Background

### 2.1  DSLs in a Nutshell

According to Fowler a *domain-specific languages* (DSL) is "a computer programming language of limited expressiveness focused on a particular domain" [17], highlighting that DSLs are typically small and only contain few domain concepts. Consider the domain of state machines used to model finite control loops in embedded systems,

■ **Listing 1** Grammar of a state machine DSL in EBNF

```
1  StateMachine = { State } { Transition } "initial" StateID ;
2  State        = ["final"] "state" StateID ;
3  Transition   = StateID "->" StateID ":" EventLabel ;
```

■ **Listing 2** Example state machine specified with the state machine DSL

```
1  state A      state B        state C         state D    // States
2  state G      final state E  final state F               // Transitions
3  A->F:1  A->B:0  B->C:1  B->D:0  C->E:1  C->E:0          // Transitions
4  E->B:1  E->D:0  F->G:1  F->A:0  G->G:1  G->G:0
5  initial A                                               // Initial state
```

of which an illustrative example is shown in figure 2 (for the moment ignoring the additional SCC information). In this domain, there are only states linked by directed transitions. Additionally, while one state must be marked as the initial state, multiple states can be denoted final state. Finally, both states and transitions have a label denoting a name and a triggering event, respectively.

Now, to create a DSL for state machines, a corresponding grammar needs to be defined. A possible grammar employing the *extended Backus-Naur form* (EBNF) specifying the syntax of a state machine DSL is shown in listing 1. It declares three rules each declaring a nonterminal (left-hand side) and productions for each nonterminal (right-hand side). Productions, in turn, can contain terminals ( `"initial"` ), optional elements ( `["final"]` ), as well as repeating elements ( `{ State }` ). For simplicity, `StateID` and `EventLabel` denote valid identifiers. In sum, this grammar defines the concrete syntax for state machines specified in this DSL.

Fortunately, most language workbenches can automatically generate a parser and syntax highlighting editor for the given grammar [15]. Assuming such an editor was created, domain experts can now exactly specify the state machine from figure 2 with the specification of the state machine DSL, outlined in listing 2. While this allows domain experts to define state machines, such that they are both human- and machine-readable, machines usually operate on a DSL's abstract syntax. There, all unnecessary terminals are removed and relevant information is retained in nonterminals and their children. This notion will be picked up in section 2.3.

## 2.2 Static Analysis

While DSLs can improve communication and productivity of domain experts [17], their ability to support static analysis tasks is often neglected. These range from declaration-use analysis to complex validations of program properties and constraints.

### 2.2.1 Cycle Detection in State Machines

In case of the state machine DSL, experts may want to detect cycles in a state machine, i.e., a state that is reachable from itself. The analysis can be simplified by considering the *sets* of states that are mutually reachable. Every such set of states forms a *strongly*

■ **Algorithm 1** Algorithm for computing strongly connected components in directed graphs, adapted from Kosaraju and Sharir [1, 38]

---

**input** : A directed graph $G = (V, E)$
**output** : A set of SCCs, whereas each SCC is represented by its set of vertices.

1   Create an empty map $A$ from vertex to SCC;
2   Create an empty list $L$ of vertices;
3   **foreach** *vertex $v$ in $V$* **do**
4     **if** *$v$ was not yet visited* **then**
5       Perform a post-order depth-first-traversal in $G$ starting from $v$, prepending each newly encountered vertex $w \in V$ to $L$;

6   Construct the inverted graph $G_r = (V, E_r)$ where, i.e., $E_r = \{(b, a) | (a, b) \in E\}$;
7   **foreach** *vertex $w$ in $L$* **do**
8     **if** *$w$ not yet assigned to an SCC in $A$* **then**
9       Create a fresh SCC $scc_i$;
10      Perform a depth-first-traversal in $G_r$ starting from $w$, assigning each unassigned vertex $v$ to $scc_i$ in $A$;

11   **return** the set of all SCCs;

---

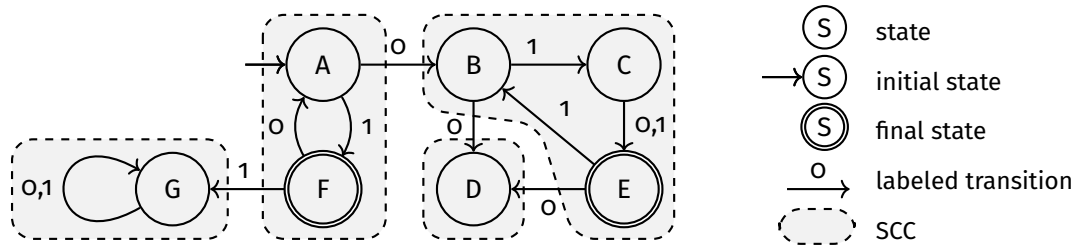

■ **Figure 2** State machine with strongly connected components

*connected component (SCC)* [38]. An efficient method for computing SCCs based on depth-first-traversals was first introduced by Kosarju and later by Sharir [38]. Accordingly, performing the algorithm shown in line 11 for the state machine example results in four SCCs, shown in figure 2.

After computing the set of SCCs, we need to give feedback to the language's user. Considering state machines, this includes filtering out trivial SCCs, which do not encompass a cycle, such as the SCC only containing state D. As a result, the algorithm detects three cycles in the example. This information can be valuable for a domain expert and helps refactoring the state machine.

### 2.2.2   Variable Shadowing Analysis

Another example of a static analysis is the shadowing of names in (mostly) hierarchically structured scopes. Shadowing describes the denial of (direct) access to a declared entity by a reference when another declaration of the same name is found earlier during the name resolution process. This can lead to various hard-to-find programming errors when names are inadvertently reused [30, pp. 132]. While some cases of shadowing are prohibited in specific languages, it is still allowed in other

■ **Listing 3**   Several (permitted yet discouraged) variable shadowings in two Java classes

```java
public class A {
  protected int x = 1;      // declare field x₁
  public A(int x) {         // constructor parameter x₂, shadowing field x₁
    this.x = x;             // use of 'this' to write in shadowed field x₁
  }
  void m() {
    int x = 3;              // declare local variable x₃, shadowing field x₁
  }
}
public class B extends A {
  int x = 4;                // declare field x₄, shadowing field x₁
  class C {
    private int x = 5;      // declare field x₅, shadowing field x₄
  }
}
```

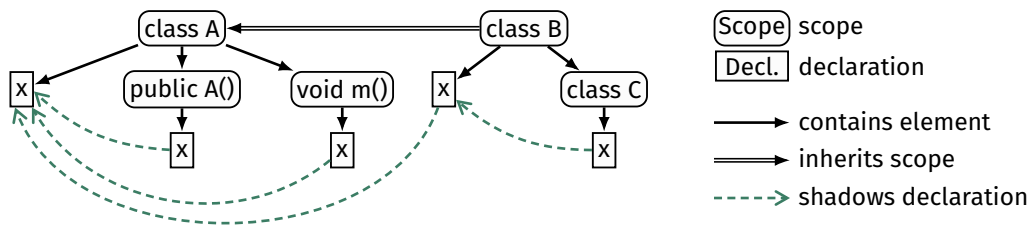

■ **Figure 3**   Scope tree with shadowing relations for listing 3

cases. Here, two examples are investigated further: shadowing of fields by other fields and local variables in Java and shadowing of variables in Modelica.

Listing 3 shows a simple Java example exhibiting shadowing of fields by variables, method parameters and other fields permitted by the language. The corresponding scope tree is shown in figure 3. A simple tree is not sufficient for a scope analysis, since fields inherited by superclasses can also be shadowed.[1] Thus, a declaration-scope data structure must contain another relation to represent inheritance.

"Modelica is a free object-oriented modeling language with a textual definition to describe physical systems in a convenient way by differential, algebraic and discrete equations." [4] In particular, the main language feature for structuring the equations is the class concept, of which many other structures are specializations of (e.g., `model`). Since Modelica classes support (multiple) inheritance and nesting, shadowing can happen, as shown in listing 4.

In conclusion, variable shadowing is an interesting analysis for three reasons. First, it can lead to hard-to-detect errors and, secondly, it occurs in various different languages.

---

[1] Additionally, visibility must be considered. This can be done by nesting multiple subscopes in a class-scope containing field declarations according to their visibility levels; then instances of an inheritance relation must point to the correct nested scope.

[2] Abbreviated version of test from `github.com/modelica/Modelica-Compliance/`

■ **Listing 4**  Example of an illegal variable shadowing in Modelica[2]

```
1  model EnclosingClassLookupShadowedConstant
2    constant Real x = 4.0; // declare constant variable x₁
3    model A
4      Real x = 3.0;        // declare variable x₂, shadowing x₁ in line 2
5      model B
6        Real y = x;        // refers to x₂ in line 4 illegally, since references
7      end B;               // to enclosing scopes must be constant [4, §5.3.1]
8      B b;
9    end A;
10   A a;
11 end EnclosingClassLookupShadowedConstant;
```

Finally, since shadowing is not in all cases prohibited, many compilers do not utter warnings, thus requiring additional static analysis tools.

### 2.3  Reference Attribute Grammars

While the preceding sections discussed DSLs and static analysis on them on a theoretical level, this section focuses on the practical aspects of implementing a DSL and corresponding static analysis. In particular, we focus on *Reference Attribute Grammars* (RAGs) as implementation technique.

*Attribute grammars* [24] are a concept to specify computable *attributes* of nodes in derivation trees (ASTs) of context free grammars. These attributes exist in addition to the *intrinsic* attributes that grant the access to the tree's tokens. Originally, two kinds of computable attributes have been proposed, *synthesized* attributes, which are computed using the children of the given node, and *inherited* attributes, which use the node's ancestors. There are many extensions and specializations of this concept, some of which we will discuss here.

*Reference attribute grammars* [20] permit values of both intrinsic and computed attributes to be references to other nodes in the tree in addition to values. Essentially, these *reference attributes* compute an *overlay graph* over the tree and thus facilitate the specification of typical attributes, e.g., for name and type analysis. *Higher order attribute grammars* [42] allow attributes to compute new subtrees that are integrated into the original tree and thus attributed and evaluated like the rest of the tree.

*JastAdd* [21] is a RAG system that uses a Java-based DSL to specify attributes and compiles the grammar specification including the attributes into plain Java. Benefits of *JastAdd* are a lazy, memoized, and incremental attribute evaluation as well as a very modular and extensible specification language using concepts of aspect-oriented programming for both grammar and attributes. This section describes relevant elements of the *JastAdd* grammar and attribute specification.

The *JastAdd* grammar specification uses a modified EBNF syntax. As an example, listing 5 shows a *JastAdd* grammar for the state machine in listing 1, which is a slightly modified version of the grammar presented in [19]. Repetitions are specified with a `*`, optional nodes are put in square brackets `[]`, and terminal symbols are placed in angle brackets `<>`. Child productions may have a context name prefix, separated

■ **Listing 5** Grammar of a state machine DSL in JastAdd notation

```
1 StateMachine ::= State* Transition* <InitialStateID:String>;
2 State        ::= <ID:String> <Final:boolean>;
3 Transition   ::= <ID:String> <FromStateID:String> <ToStateID:String>;
```

with a colon. The alternative rule is modelled with rule inheritance, in addition, rules can be abstract. The first introduced nonterminal, *StateMachine*, defines the only start symbol as it does not occur on the right-hand side in any of the rules, however, there is no explicit annotation to mark the start symbol, thus, there may be several potential start symbols in a grammar.

**Synthesized attributes** are declared with the `syn` keyword and compute their value using child nodes. The definition of the attribute equation starts with `eq`, but may also be attached to the declaration if it is defined on the same type.

```
1 syn boolean StateMachine.numberOfElements();
2 eq StateMachine.numberOfElements() = getNumState() + getNumTransition();
3 syn boolean State.isFinal() = getFinalState();
```

**Inherited attributes** are computed using an ancestor of the node they are defined on. For each possible derivation tree, there must be a definition for the attribute in an ancestor of the attributed node, which is also specific to the *context*, e.g., `getState()`.

```
1 inh StateMachine State.containingMachine();
2 eq  StateMachine.getState().containingMachine() = this;
```

**Collection attributes** offer a declarative way to collect values from (parts of) the tree. Using them, no manual traversal of the tree is required.

```
1 coll Set<String> StateMachine.finalStates() [new HashSet()] with add;
2 State contributes getID() when isFinal() to StateMachine.finalStates();
```

**Reference attributes** are attributes that return references to other nonterminals of the AST. They can take the shape of all aforementioned types of attributes.

```
1 syn State StateMachine.initial() {
2   for (State state: getStateList())
3     if (state.getID().equals(getInitialStateID()))
4       return state;
5   return null;
6 }
```

**Higher order attributes *or* nonterminal attributes (NTA)** return nonterminals just like reference attributes. However, they create *new nonterminals* rather than creating a *new reference* to an existing node. After creation, the root of the new subtree is integrated into the tree just like a regular child node and can be analysed and navigated as such by attributes, including its intrinsic *parent* attribute.

The following example extends the grammar with a new nonterminal *Metadata* and adds a computed metadata element to each state.

```
1 Metadata ::= <Label:String> <Final:boolean> ;
```

```
1 syn nta Metadata State.getMetadata() = new Metadata(getLabel(), isFinalState());
```

Using a set of grammar rules and attribute specifications defined in extensible and refinable *aspects*, *JastAdd* constructs a system to analyse trees. Hereby, it uses lazy, demand-driven attribute evaluation, the memoization of attribute values, and still allows the modification of the tree and a subsequent incremental attribute evaluation by employing dynamic dependency tracking [13, 40].

RAGs can be employed to develop and analyse DSLs. However, *computed* reference attributes show a problem: since all references are denoted by identifiers of the referred states, it is easy to introduce inconsistencies. Therefore, the following section introduces relations, a concept to enhance the support of intrinsic references to RAGs.

## 3 Static Analysis with Relational RAGs

### 3.1 Relational RAGs

Recently, the concept of RAGs has been extended to support *non-containment* relations [32, 33], which has proven beneficial when describing conceptual models with attribute grammars. In the following, we call this extension *relational RAGs*. In RAGs, references are typically resolved by evaluating reference attributes. However, this incurs inconsistencies when references can not be resolved, and either inconsistencies or additional computational effort when bidirectional relations should be modelled.

In *JastAdd*, a relation is added to the grammar specification with the `rel` keyword followed by a pair of annotated types and a direction (`->` , `<-`, or `<->` ). While a relation does not have a name, much like named contained children, each outgoing type of a relation has a *role name* and a multiplicity of `?` , `*` , or (by default) one.

■ **Listing 6** Grammar of a state machine DSL

```
1 StateMachine ::= Element*;
2 abstract Element ::= <Label:String>;
3 State : Element;
4 Transition : Element;
5 rel Transition.From <-> State.Outgoing*;
6 rel Transition.To <-> State.Incoming*;
7 rel StateMachine.Initial -> State;
8 rel StateMachine.Final* -> State;
```

Listing 6 shows a variant of the state machine grammar with relations instead of references computed by name lookup. The relations at lines 5 and 6 are bidirectional relations replacing the names in the production rule at line 3 in listing 1 while the unidirectional relations at lines 7 and 8 replace the *Final* flag and the *InitialStateID* in the original grammar. Thus, the addition of relations to RAGs simplifies and enhances the description and traversal on non-containment relations *within* an AST. In the

■ **Listing 7**  Attributes for cycle detection attributes defined for a state machine

```
1  syn Set<Set<State>> StateMachine.SCC() {
2    Map<State, Set>  visited = new HashMap<>();
3    LinkedList<State> locked = new LinkedList<>();
4    for (State n : states())
5      if (!visited.containsKey(n))
6        n.visit(visited, locked);              // forward search
7    for (State n : locked)
8      if (visited.get(n) == null)
9        n.assign(visited, new HashSet());      // backward search
10   return new HashSet(visited.values());
11 }
12 void State.visit(Map<State, Set> visited, LinkedList<State> locked) {
13   visited.put(this, null);
14   for (Transition t : getOutgoingList())
15     if (!visited.containsKey(t.getTo()))
16       t.getTo().visit(visited, locked);
17   locked.addFirst(this);
18 }
19 void State.assign(Map<State, Set> visited, Set root) {
20   root.add(this);
21   visited.put(this, root);
22   for (Transition t : getIncomingList())
23     if (visited.get(t.getFrom()) == null)
24       t.getFrom().assign(visited, root);
25 }
```

following sections, we use the two examples from section 2.2 to demonstrate how relational RAGs can help with the definition of static analysis.

### 3.2 Static Analysis with Relational RAGs: Detecting Cycles in State Machines

Static analysis is a frequent application for RAGs, e.g., for control and data flow analysis [39] or null-checks [12]. Especially the idea to attach both the analysis algorithms *and* the analysed, computed properties of a derivation tree directly to the corresponding nodes in a formally prescribed way enables the concise specification of static semantic properties of an AST. Additionally, the *JastAdd* RAG tool further improves the suitability by allowing an aspect-oriented attribute specification and providing an attribute specification DSL embedded in Java to provide a familiar programming language for it.

The implementation of Kosaraju's algorithm (line 11) with *JastAdd* as shown in listing 7 is straightforward. The input data structure is a state machine AST, thus, vertices $v \in V$ are *State*s and edges $e \in E$ are *Transition*s. The data structures $A$ and $L$ are implemented using a `Map` and a `LinkedList`, respectively, whereas the result is returned in a `Set` of `Set` s. Since the two depth-first traversals are best described recursively, the attribute employs two helper functions defined in *State*, `void visit()` and `void assign()` in lines 12 and 19. Note that these methods are *not* attributes, since they modify the content of the lists passed to them as arguments, thereby violating the rule that attributes must be side-effect free. Furthermore, observe that

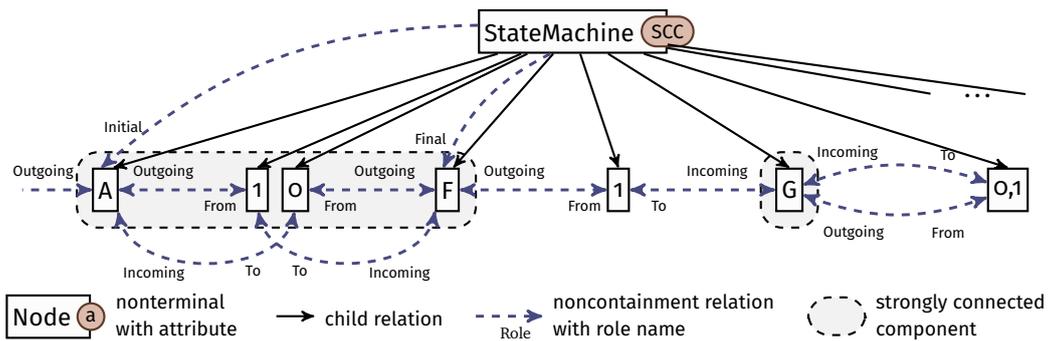

■ **Figure 4** Derivation tree of the *StateMachine* grammar with non-containment edges the `visit()` method uses the forward-direction (`getOutgoingList()` and `getTo()`), while `assign()` moves backwards (`getIncomingList()` and `getFrom()`).

Figure 4 shows parts of the derivation tree of the state machine from figure 2. Two types of edges can be distinguished: *containment* edges from a parent to its children form the tree while the dashed edges represent the non-containment relations defined by the relational RAG. Considering only the relations between states and transitions defined in lines 5 and 6 of the grammar in listing 6, two things can be observed. First, both relations are bidirectional, enabling the direct application of Kosaraju's algorithm. Secondly, the edges $E$ and $E_r$ of the graph used in the algorithm are *not* instances of the relations in the grammar, which have an additional *Transition* nonterminal.

While this is a concise and efficient implementation of line 11, it can only be applied to one specific data structure, the *StateMachine*. Not only is the attribute *SCC* defined for this non-terminal, but the result as well as the internal variables and helper methods use the API derived from the state machine grammar. Furthermore, the algorithm requires the graph to be navigable in both directions, which is possible in this particular example, but certainly not in all structures representing graphs. Therefore, a more reusable approach to specify static analysis is needed and presented in the next section, using relations *between* trees rather than just within a single tree.

### 3.3 Static Analysis with Relational RAGs: Detecting Variable Shadowing

The second introduced example – shadow analysis – is even better suited for relational RAGs, because, as figure 3 shows, the main data structure on which the analysis is performed is a tree.

- First, each element can easily be assigned to a containing scope with an inherited attribute: **inh** `ASTNode ASTNode.containingScope();`
- Then, the declarations can be collected, using the previous attribute to determine the scope: **coll** `HashSet<Declarator> ASTNode.elementsInScope();`
- Using the scopes and their contents, an attribute can be defined to find a shadowing declaration, if one exists: **inh** `Optional<Declarator> Declarator.findShadower();`
- Finally, all shadowed declarations can be collected:
  **coll** `HashSet<ShadowFinding> ASTNode.shadowedDeclarations();`

While this approach sounds straightforward in theory, there are some caveats. First, it must be ensured that all cases are covered: all language constructs that specify a scope must be considered and the correct scope must be assigned for each declaration (which may not always be *syntactically* contained). Secondly, visibilities and inheritance must be considered. In conclusion, two stages can be observed. First, the collection and structuring stage, which collects declarations into nested scopes. Second, the analysis stage, where the analysis algorithm is executed. While RAG-based analysis permits a separated specification of the stages, execution is interleaved, complicating debugging of both stages. Furthermore, while the collection stage is very grammar-specific, the analysis does not have to be. Again, this suggests benefits from employing a more reusable analysis.

## 4 Relational RAGs for Reusable Static Analysis

The RAG-based analysis presented in the last section is short and efficient. However, it is difficult to reuse, since in about half of the lines in listing 7, concepts of the underlying grammar are referenced. Even though the presented algorithm only serves as a small example for a large set of potentially much larger and more complex static analysis algorithms, already in this case a copy-and-paste reuse with subsequent modification is error-prone, mostly because of the model navigation entangled in the algorithm. Thus, this section first introduces more use cases that could benefit from the same kind of analysis and presents means to make it reusable.

### 4.1 A Case for Reusable Static Analysis

Revisiting the presented analyses in section 2.2, two observations can be made.

**Reuse**    The analyses themselves work on rather simple data structures, a directed graph and a tree with additional relations. This implies that if there would exist a transformation into these data structures, the analyses could be reused. Furthermore, there actually is need for reuse – both analyses can be applied to multiple languages. In the case of cycle detection, a very common example of cycles that are analysed are dependencies between components, e.g., dependencies between classes in object-oriented programs. This is an important software design issue, as, e.g., Lakos acknowledges: "Although we might be serene enough to tolerate cyclic dependencies among a few components within a single package due to carelessness, ignorance, or special circumstance, we must be steadfast in our resolve to avoid cyclic dependencies among packages." [28, page 496]. Thus, we reuse the presented dependency analysis with two kinds of dependencies in Java programs, *type* and *package* dependencies, showing not only reuse in different languages, but also within one language.

**Separation of Concerns**    As mentioned in section 2.2.2, separating the gathering of information, i.e., the traversal of the AST, from the actual analysis is also very beneficial

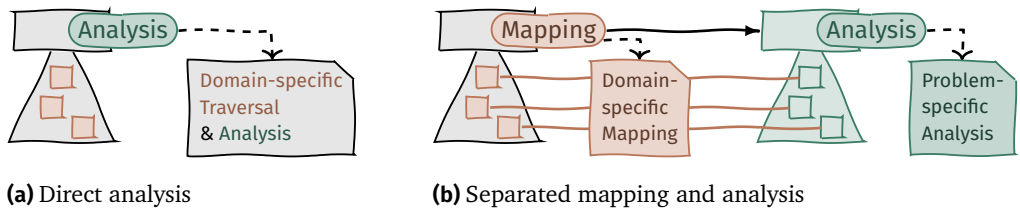

**(a)** Direct analysis       **(b)** Separated mapping and analysis

■ **Figure 5**   Direct and decoupled information gathering and analysis for debugging. An explicitly defined, navigable data structure can be analysed and printed, e.g., to determine whether all cases have been covered.

Figure 5 contrasts the direct approach with the split approach. Simultaneously, figure 5b indicates the required parts: a problem-specific data structure, mapping relations and a mapping attribute, and a problem-specific analysis. In the following, a strategy to create a reusable analysis by specifying these components is presented and applied to the cycle detection analysis. Subsequently, differences and similarities in the application to the shadowing use case are briefly discussed.

## 4.2 Externalizing Static Analysis

In section 4.1 it was already mentioned that to externalize static analysis, four components are required. Now that the prerequisites have been discussed, these components can be introduced along with a process to create them.

**Define a Problem-Specific Grammar** This grammar must contain *all* required information to efficiently perform the analysis. Even though references into the domain-specific tree are added in a later step, they can not be used during analysis as this would impede its reuse for another language. Considering the requirements of the analysis, directions of noncontainment relations must be selected accordingly.

**Specify a Static Analysis Attribute** Using the grammar defined in the first step, the analysis attribute can be written. During this process, missing information in the grammar may be detected that require another iteration of the previous step.

**Define the Mapping Relations** These domain-specific relations are defined in a grammar that (only) contains relations from the problem- to the domain-specific types.

**Construct the Mapping Attribute** The mapping attribute derives the problem-specific tree and contains instances of the relations defined in the previous step.

The following sections discuss technical requirements for this analysis and outline aspects of the process relevant to the examples. Afterwards, section 5 illustrates the process by implementing the example analyses using *JastAdd*.

## 4.3 Decoupling Strategies with Relational RAGs

The presented approach introduces a mapping between two trees, the domain-specific and the newly created problem-specific tree. Before the process to externalize analysis can be sketched, the means of creating and maintaining inter-tree relations must be

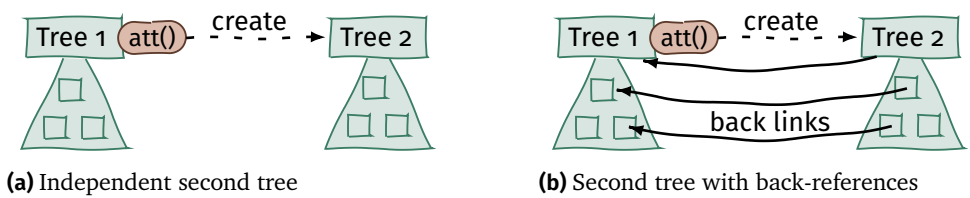

**(a)** Independent second tree

**(b)** Second tree with back-references

■ **Figure 6** Attribute-based tree construction with relations discussed. To relate multiple trees to one another, both the nature of the relation, i.e., its directionality, and how the relation is constructed have to be considered. Since we focus on the specification of reusable analysis algorithms and structures, one of the trees should contain information *synthesized* from the existing derivation tree. While relational RAGs presented in section 3.1 already provide relations to connect elements in different trees, they do not prescribe how to obtain them. Thus, to understand how a new tree can be constructed based on an existing one, the means to construct and modify trees are presented.

**Deriving new Trees from Existing Trees**    In principle, an attribute grammar expects an already existing, immutable tree that has been constructed beforehand, typically using a parser. However, nonterminal attributes (NTAs) are an approach to construct subtrees using attribute evaluation, a process that in *JastAdd* is transparent to any other attribute evaluation and tree traversal, because it is performed on-demand when the NTA is first encountered by an attribute. If a *new* tree instead of a subtree should be constructed, regular attributes that happen to return newly created subtrees can be used as long as they are memoized.[3]

The construction of a derived tree using an attribute has some consequences for how it can be used. First, an attribute-based construction builds exactly *one* new tree for every base tree, since, by definition, attributes have one value for any given tree.[4] Secondly, while it is possible to modify elements in the base tree by employing attribute dependency tracking, the constructed tree must not be modified. On the other hand, when nodes in the underlying tree change that the computed tree depends upon, the constructing attribute has to be re-evaluated, resulting in a whole new tree consistent with the modified base tree.

While this far, only the construction of the new tree, and thus the navigation from the base tree into it via the constructing attribute itself have been considered, there may exist further references between the trees, discussed in the next section.

**Linking Multiple Trees with Relational RAGs**    When a second tree is created using an attribute, there is a natural navigable connection to it from the original tree via the attribute, as shown in figure 6a. Additionally, there may be computed reference edges

---

[3] If NTAs are not memoized, the construction is repeated, thus breaking reference attributes and relations.

[4] A second invocation would simply return the cached tree from the initial invocation.

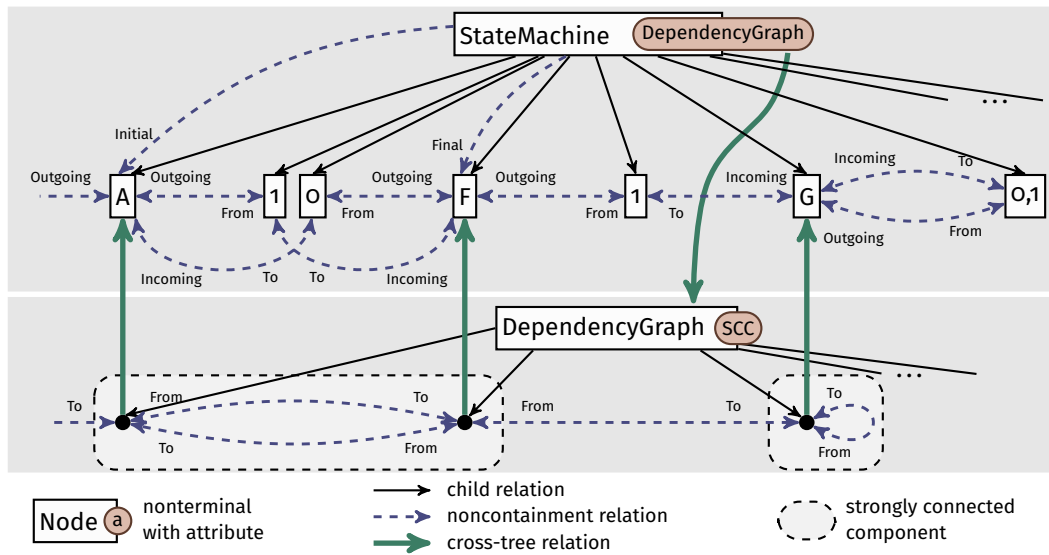

■ **Figure 7** State machine and dependency tree, linked with an NTA and cross-tree relations from the original to the derived tree, constructed in the normal way by exploiting the one existing direction.

Of course, within both the original tree and the derived tree there can be relations. To understand which relations can exist *between* trees, it has to be understood how relations are added to RAGs. Relations are a combination of intrinsic references between nodes plus synchronization mechanisms. In particular, a bidirectional relation comprises two intrinsic references, one in each direction. Thus, adding a bidirectional relation modifies the tree in *two* places, on both end points. This, however, means that connecting an NTA to the base tree involves modifying the tree in two places and in particular also directly in the base tree, which is forbidden, because it modifies the tree during the evaluation of an attribute. Thus, there may only be unidirectional relations from the derived tree into the original tree, as shown in figure 6b. The following section illustrates how the presented relations are used in the running examples.

### 4.4 Externalized Analysis for the Case Studies

Figure 7 extends the state machine tree shown in figure 4 with a derived data structure for a simple directed graph. Besides the nonterminal attribute *DependencyGraph* computing the dependency graph, the state machine is completely agnostic of it. In the shown excerpt, there are three cross-tree relations, connecting the *component* nodes of the dependency graph to the *State* nodes of the state machine. The relation between components in the dependency graph (line 3 in listing 8) are *not* connected to the *Transition* nodes of the state machine. Not only is this impossible with the proposed model, since there are no direct means to use relations as relation endpoints (or hyperedges), but it is also not required, since the labelling of the transitions is irrelevant for dependency analysis.

For the second use case, the scope tree shown in figure 3 already provides a good idea of what the problem-specific grammar looks like. Observe that the tree with noncontainment relations between inheriting scopes perfectly match the abilities of relational RAGs.

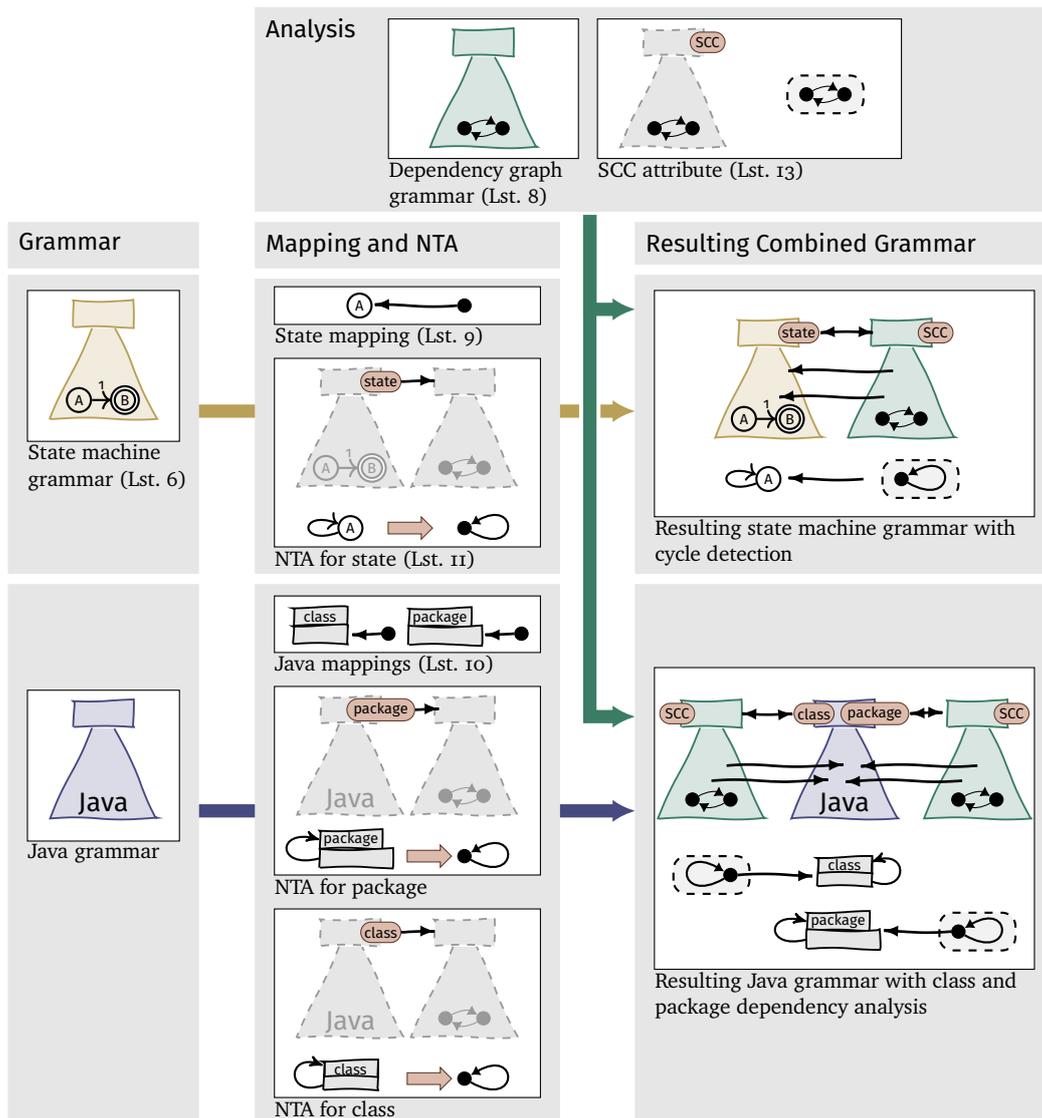

**Figure 8** Horizontal and vertical reuse of static analysis tasks

Since so far only a concept to derive new trees to perform static analysis on has been described, the following section shows how the implementation of relational RAGs can be employed in multi-tree settings, which artefacts have to be specified, and in particular how to actually perform the presented analysis.

## 5 Implementing Reusable Analysis with JastAdd

Thus far, the discussion focused on the grammar level, conversely, this section highlights the implementation of the various artefacts. Figure 8 illustrates how the cycle detection algorithm is applied to both state machines (cf. sections 2 and 3.2) and Java

■ **Listing 8** The dependency graph grammar

```
1  DependencyGraph ::= Component*;
2  Component;
3  rel Component.From* <-> Component.To*;
```

dependency analysis (cf. section 4.1) by enriching the DSLs' grammar with the analysis grammar and attributes. The upper row indicates the reusable analysis, e.g., the dependency graph grammar and the SCC algorithm as attribute. By contrast, the first column encompasses the different DSLs and their grammar. Consequently, each row depicts the artefacts required for reusing the dependency graph and cycle detection, i.e., a mapping grammar and corresponding mapping NTA; whereas the resulting combined grammar is shown in the last column.

### 5.1 Specifying the Base Language Grammar

Initially, the DSL's grammar must be specified (cf. section 2.3). While the state machine's grammar was introduced in listing 6, we reused both the Java grammar specified within the extensible compiler *ExtendJ* [13] (for Java 8) and the Modelica grammar specified in *JModelica* [3]. Notably, *ExtendJ* permits both adding analysis aspects and extending the Java grammar itself. This enabled use to implement both a baseline and reusable variants of the dependency analysis tasks.

### 5.2 Implementing Problem-Specific Analysis

To reuse a static analysis, a problem-specific, domain-independent data structure has to be designed. For cycle detection, this could be the dependency graph grammar, shown in listing 8, modeling a directed graph, which can be traversed in both directions by means of the roles `From` and `To`.

Listing 8 shows a grammar for a directed graph as used in line 11. This grammar is the minimally required grammar with a root, a list of components without any properties and a bidirectional relation between the components. The bidirectionality of the relation is a requirement of line 11, which navigates both directions.

This, in turn, obviates the need to construct the inverse graph within line 11. Besides this optimization, its implementation, as shown in the appendix in listing 13, corresponds to the algorithm. Moreover, when compared to the direct implementation, in listing 7, this algorithm avoids all domain-dependent types and navigation ensuring its general reusablity.

### 5.3 Composing Relational RAGs

Next, both the domain-dependent base language and the analysis' data structures need to be composed. Fortunately, both *JastAdd* and its relational extension support grammar and attribute composition. Since these grammars encompass production rules and relation definitions, their composition is obtained by collecting all rules

■ **Listing 9** The mapping relations for the state machine example

```
1  rel Component.State -> State;
2  rel DependencyGraph.StateMachine -> StateMachine;
```

■ **Listing 10** The mapping relations for both Java examples

```
1  rel DependencyGraph.Program -> Program;
2  TypeComponent : Component;                    // grammar and
3  rel TypeComponent.TypeDecl -> TypeDecl; // relations for type analysis
4  PackageComponent : Component ::= <Package:String>; // package extension
```

and relations. As a result, a grammar can be extended by means of nonterminal inheritance and relation definition. The composition of attributes in *JastAdd* is more sophisticated. Besides the collection of attribute declarations and equations in *aspects*, aspect-oriented techniques, such as method refinement and wrapping are supported.

**Defining Mapping Relations** To map the domain-specific concepts to the analysis' domain-independent constructs, a grammar extension must be defined. This extension only defines relations from elements of the analysis tree back to the base tree.

Listing 9 shows the extension of the state machine DSL. The first relation links a *Component* to a *State*, whereas the second links the roots of both trees. The latter effectively makes the computed nonterminal relation bidirectional, where the opposite direction is represented by the (computed) NTA.

In case the same analysis should be performed w.r.t. to different aspects of the DSL, e.g., distinguishing between class and package dependencies, a more complex mapping is required. In particular, listing 10 shows the combination of the analysis of both package and class dependencies. By employing inheritance, two variants of the dependency graph are created, one with `TypeComponents` and one with `PackageComponents` for reflecting the class respectively package dependencies. This, additionally, illustrates reusability *within* one DSL.

Aside from that, listing 10 features a corner case, where a domain-independent concept is not linked back to its domain-dependent counterpart. This is because, Java has no dedicated entity for packages. Thus, instead of a relation, it suffices to add the package name to the *PackageComponent*. Granted, the mapping depends on the DSL and use case of the reused analysis task.

**Defining the Transformation Attribute** Finally, an NTA constructing the domain-independent data structures from the base grammar must be defined. This is the most complex step and will be demonstrated for the state machine DSL and Java, outlined in listing 11 and listing 14, respectively.

For the former, the relational NTA involves three steps. First, a new tree is initialized and related to the base tree. Second, for each state, a component is created and linked back to the corresponding concept in the base grammar. Since this relation cannot be bidirectional, a mapping must be stored locally within the attribute (cf. line 4).

■ **Listing 11** The relational NTA to compute the state dependency graph

```
1  syn lazy DependencyGraph StateMachine.dependencyGraph() {
2    DependencyGraph dg = new DependencyGraph();
3    dg.setStateMachine(this);
4    Map<State,Component> componentMap = new HashMap<>();
5    for (State s: states()) {
6      Component n = new Component();
7      n.setState(s);
8      dg.addComponent(n);
9      componentMap.put(s, n);
10   }
11   for (Transition t: transitions())
12     componentMap.get(t.getFrom()).addTo(componentMap.get(t.getTo()));
13   return dg;
14 }
```

Finally, all transitions are transformed into component dependencies by using the local mapping. Note that the `lazy` modifier enables memoization for the relational NTA and ensures the validity of references and attribute values.

The implementation is straightforward and will be similar for most DSLs, as both components and dependencies are collected in two separate loops. For instance, the relational NTA for the Java class dependency graph, shown in the appendix in listing 14, has a similar structure. The main difference here, are the traversals of the Java grammar defined by means of two collection attributes. The attribute `Program.typeDecls()` collects all components, while `TypeDecl.typeUses()` collects all uses of types within a type definition, and thus its dependencies.

**A Template for Transformations** Thus far, we described the transformation attribute for one particular use case. In general, the implementation of transformations, especially for large and complex languages, can be structured using a pattern. Henceforth, we illustrate this pattern by outlining the implementation of the shadowing analysis for Modelica. While the source code for this analysis can be found in the appendix A.2, listing 12 illustrates the required parts of the transformation, separated into *JastAdd* aspects.

- `ModelicaToScopeMapping` contains the actual mapping attributes that perform the tree traversal and collecting the result in a scope tree that is returned. This aspect uses the attributes defined in the following two aspects.
- `MappingConstructors` constructs the individual nodes of the scope tree. These mappings currently have to be defined manually, yet they could also be generated automatically from the mapping relations.
- `ScopeGenerationAttributes` contain the remaining helper attributes required to perform the analysis. In case of the scope analysis, this is primarily the inherited `containingScope()` attribute.

In conclusion, there is work required to define the transformation into a domain-independent structure, but it is feasible even for a large base grammar. While the presented case studies alone can not show an overall reduction in total effort, particu-

■ **Listing 12** Transformation attributes from Modelica to a scope tree

```
 1  aspect ModelicaToScopeMapping {
 2    syn lazy RootScope SourceRoot.scopeTree() {
 3                    // invoke constructors from aspect 'MappingConstructors'
 4      return tree; // and attributes from aspect 'ScopeGenerationAttributes'
 5    }
 6    // more relational nta attributes and helper methods
 7  }
 8  aspect MappingConstructors {
 9    // rel ClassDeclScope.classDecl -> SrcClassDecl;
10    syn lazy ClassDeclScope SrcClassDecl.asScope() {
11      ClassDeclScope scope = new ClassDeclScope();
12      scope.setClassDecl(this);
13      return scope;
14    }
15    // more constructors
16  }
17  aspect ScopeGenerationAttributes { /* helper attributes */ }
```

larly because the SCC analysis implemented in listing 13 and the shadowing analysis is quite concise, even in these examples there are reasons why creating a reusable analysis is beneficial.

First, for a developer familiar with the base language grammar, the construction of the glue structures is a simple task, because, by definition, the analysis structures are as small as possible and the transformation algorithm thus does not entail much *additional* overhead compared to a direct implementation. Secondly, the separation of concerns of grammar traversal and analysis enables variability in the analysis – as long as it uses the same data structure, it can simply be replaced. Thirdly, the separation also improves debugging. Because the relevant structure is made *explicit*, it can be analysed easily. Additionally, the algorithm can be debugged using simpler test structures instead of ones derived from a large real-world language. This can be utilized both for finding bugs in the implementation and also performance issues.

Considering this, performance is also relevant for another reason. The presented approach adds another data structure, which itself requires both time and memory and may introduce a certain overhead during the execution of the algorithm. Thus, the approach is evaluated with respect to its performance in the following section.

## 6  Performance Evaluation

While the presented approach has benefits for creating and applying static analysis, the modular structure may have an influence on the performance of reused code. In particular, the presented approach is based on *duplicating structures*, i.e., the very structures on which the analysis is performed, exists twice – once in the domain-specific and once in the algorithm-specific form. Additionally, cross-tree relations are required to link the two representations. While this certainly leads to increased memory requirements, it also suggests a runtime overhead induced by the construction time of the additional structures. On the other hand, the analysis algorithm itself may profit from the specifically tailored data structure it operates on. This section uses the Java dependency cycle detection analysis to investigate these issues by performing the analysis on a large corpus of open source projects written in Java. The complete source code, all scripts to obtain the measurements, and the data collected from our benchmark system are available.[5]

**Experimental Setup**    Besides the artefacts mentioned in figure 8, we implemented two baseline implementations of the algorithm for type- and package-based SCC analysis. To assess the differences in performance for the presented approach, we used the *Qualitas Corpus* [41], a collection of the source code of 112 well-known open source Java projects, including large examples such as *Eclipse*, *JBoss*, and the *Hibernate* framework. For every project, each contained Java file was considered.

To evaluate the runtime behaviour, we performed each of the two analysis kinds with both the direct and the reusable variant of the analysis 101 times and measured the runtime. The experiment was run on an Intel i7-8700 workstation with 64 GB of memory using Fedora Linux 29 running on kernel 4.18, OpenJDK version 1.8 and *JastAdd* version 2.3.3.

**Measurement Results**    To give an idea of the general performance of the analysis, figure 9 shows box plots of the analysis runtime for the eight largest projects. Note that all shown times are the total analysis times, which include the runtime of the transformation attribute, but *not* the parsing time, i.e., the time to read in the file and construct the AST corresponding to the source code. The compactness of all boxes implies little variance in the runtime, supported by the fact that besides some uses of hash maps in *JastAdd* the algorithms have a deterministic control flow. We assume that the maximum runtime deviations shown by the upper whiskers are due to memory management and garbage collection of Java that is required when parsing and analysing such a large number of files.

For a more detailed look, the appendix contains tables 1 and 2 with measured numbers for the eight largest projects for package and type dependency analysis. The size of the projects is stated with both the number of Java files that were analysed and the sum of the number of its nodes and edges in the dependency graph. Additionally, the median runtime out of 101 measurements for both analyses is shown. Finally, the median overhead of the reusable analysis compared to the directly defined one is stated in the last column. For both analyses, the tables show that *JastAdd* is capable of performing both direct and reusable analysis in less than 80 s even for a very large Java project like *NetBeans* with 32 647 files. Additionally, the results show that both approaches have a very similar performance and, surprisingly, that for most of the projects the reusable analysis is actually faster, resulting in a negative overhead.

---

[5] The implementation is available at [31] and https://git-st.inf.tu-dresden.de/jastadd/reusable-analysis.

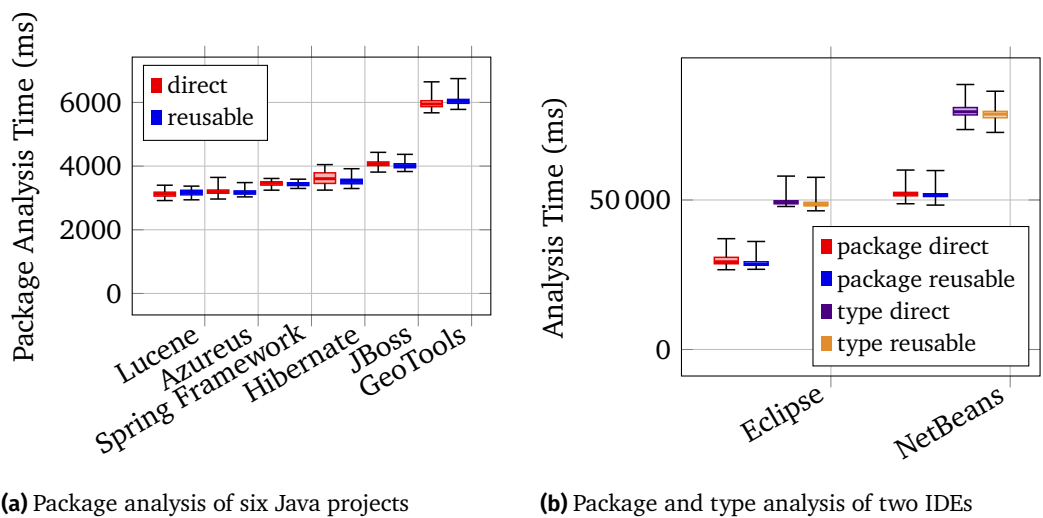

**(a)** Package analysis of six Java projects

**(b)** Package and type analysis of two IDEs

■ **Figure 9** Dependency analysis of the biggest Java projects

While the results presented so far show that reusable analysis is feasible for large projects, figure 10 shows the median analysis times for all projects. Again, figures 10a and 10c show that the direct and reusable analysis have almost the same performance. The separate box plots in figures 10b and 10d show that for the package analysis the median overhead is very small while for the type analysis the overhead is negative. In these plots, the whiskers demarcate 2.5th and the 97.5th percentile, but even the outliers are below 10 % in either direction.

The better runtime of particularly the type analysis shows a benefit of the presented approach of using a problem-specific data structure. While *JastAdd* has the powerful tool of collection attributes, which allows a concise and declarative definition of inverse direction of a relation, still a *computation* is required, whereas the bidirectionality of the type usage relation in the dependency graph has no computational and very little memory overhead. In the case of package usage relations, this benefit is not as visible, because there are far fewer packages, and thus package dependencies, so their computation is not as influential for the total runtime. The following section discusses the results of the case study and the their general implications.

## 7 Discussion

The investigated languages and analyses as well as the performance evaluation offer several insights in the presented approach for reusable analysis.

**Results** First, *JastAdd*, *ExtendJ*, and *JModelica* are viable tools to perform such an analysis and, thus, (relational) RAGs are a suitable approach. Secondly, the approach supports the specification of reusable analyses, both in different contexts within one domain, as shown by the two Java use cases, and in a completely different domain, as shown with the state machine. Finally, the large-scale benchmark performing the

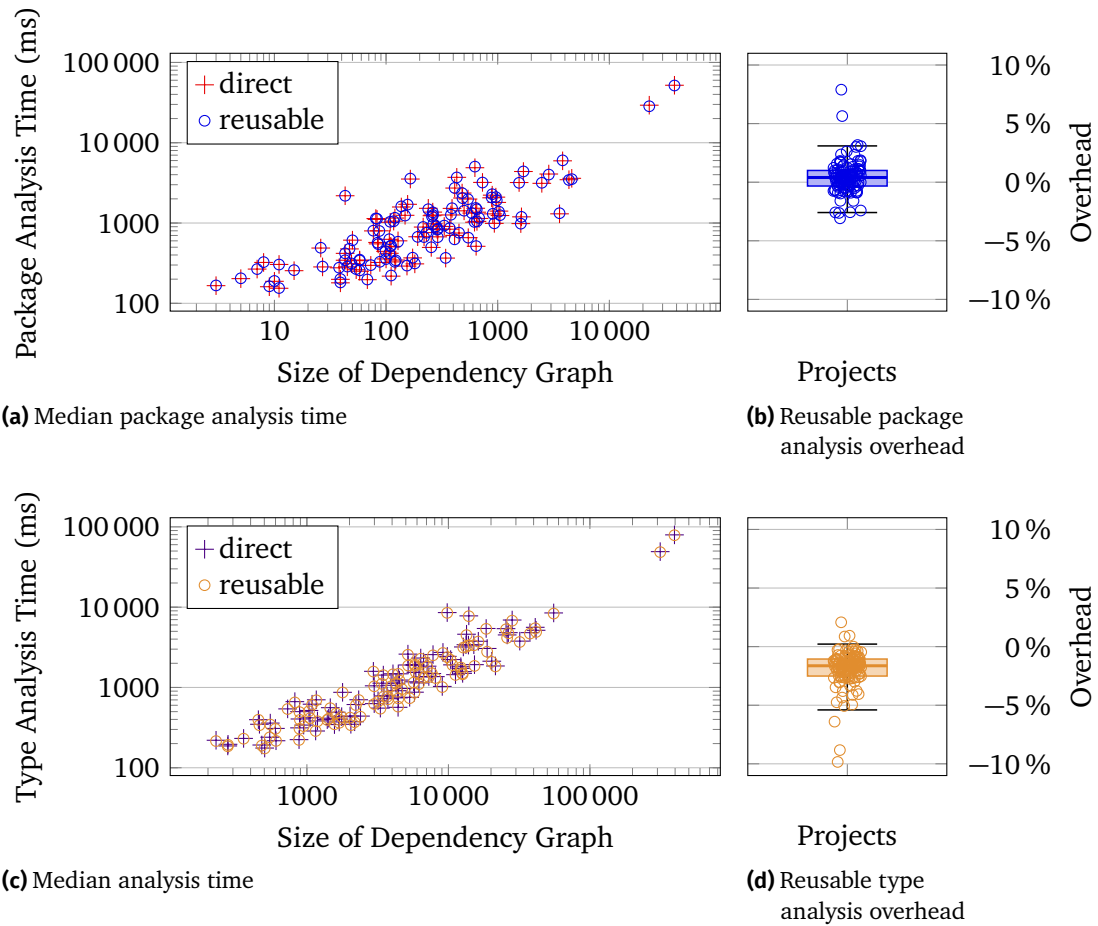

**(a)** Median package analysis time

**(b)** Reusable package analysis overhead

**(c)** Median analysis time

**(d)** Reusable type analysis overhead

■ **Figure 10** Comparison of median analysis times and accumulated overhead of package and type analysis for all projects in the Qualitas Corpus analysis on real-world programs shows a negligible overhead and in some cases even an observable speed-up when reusing analyses (cf. figures 10b and 10d).

**Limitations**   As the implementation of the presented algorithm is based on two well-known and frequently used analysis, it is very concise, so at least in terms of lines of code, the benefits of reuse are limited. We do not yet know, how well the approach works on a wide variety of analysis algorithms. Similarly, we applied it to a very simple and two very complex programming language, certainly two extremes, most DSLs rank in-between. Additionally, not all kinds of static analysis can be described concisely and efficiently using RAGs. In particular, analyses that require much modifiable state are hard to map to the tools RAGs offer. Especially, in cases where the analysis requires a very fine grained mapping of DSL elements – as is the case for control flow analysis – the implementation effort for the mapping increases compared to the analysis algorithms on a control flow graph [39].

**Opportunities**   However, the presented approach can be further extended. One promising idea could be to stack problem-specific data structures. For the cycle detection, e.g., another data structure can be derived from the dependency graph NTA that contains just the strongly connected components and their connections, which can be further processed, e.g., for displaying them. This additional step has already been defined for the dependency graph to create a representation with all required information to create *GraphViz* and *PlantUML* visualizations. Furthermore, instead of specifying the analysis on a specialized data structure, this structure could also be used to generate a specification for an existing, specialized analysis tool, such as ILP solvers [37], *interprocedural, finite, distributive subsets* (IFDS) solver [7, 11] or deductive theorem provers [2]. This permits integrating and reusing analyses that are neither already available for a given DSL nor easily and efficiently implementable using RAGs.

The next section places this work in the context of state-of-the-art and related work.

## 8  Related Work

The implementation of custom static analysis is supported by most state-of-the-art language workbenches [15]. However, when the study discusses validation it considers the following static analysis tasks: structural validation, name resolution, type checking and programmatic validation. While the former three denote typical analyses applicable for most DSLs, only the latter can be used to implement more sophisticated static analysis. Yet, according to the authors, "many language workbenches do not provide a declarative validation mechanism and instead allow validations to be implemented in a normal [general purpose language]" [15, page 28]. Nonetheless, we argue, that both declarative and procedural static analyses are tied to the DSL's underlying AST. In particular, this includes approaches that have a dedicated specification of a DSL's type system, e.g., *Spoofax* [23], *SugarJ* [14], *MPS* [43], and *Xtext* [6]. Consequently, while most language workbenches support the specification of static analysis tasks, most do not permit reusing them between different DSLs.

Besides these classic language workbenches, researchers have investigated two paths to improve reusability of static analyses. On the one hand, some language workbenches already focus on modular development of DSLs or product lines of DSLs including their static analyses, such as, *MontiCore* [26], *Melange* [10] and *Neverlang* [27]. Granted, these approaches permit modular implementation of a static analysis task having a partial implementation for each language construct. We argue that this fosters reuse on a language construct level. However, analysis tasks, like dependency analysis, cannot be reused between unrelated DSLs with different language constructs. By contrast, our approach supports exactly this kind of reuse.

On the other hand, several similar approaches specifically aim for reusable static analysis. *Boogie* [5], for instance, is a framework for static and dynamic program analysis of object-oriented DSLs that utilizes an intermediate representation, i.e., *BoogiePL*, to decouple the language syntax from the analysis task and machinery. Similar to our approach, a mapping from the DSL to *BoogiePL* must be specified, as well as feedback mechanisms. However, as *BoogiePL* is tailored to dynamic program analysis, the mapping from the DSL must be complete, regardless of whether the reused analysis actually requires all *BoogiePL* concepts or not. Notably, the *BoogiePL* can be considered a domain-independent data structure for program verification and, thus, could be reimplemented with our approach.

By contrast, the *Hoopl library* [35] and the *Galois Transformers* [9] aim at providing modular, reusable operators and transformations for program analysis that can employed to implement program analyses in Haskell, such as, dataflow analysis and abstract interpretation, respectively. The former provides a library of reusable polymorphic operators and transformation, which allow compilers to implement dataflow optimizations [35]. Similarly, *Galois Transformers* represent domain-independent monadic components for constructing and reusing program analysers [9]. Both, *Hoopl* and *Galois Transformers* show compositional soundness and correctness, yet, neglect performance which we argue is the main blocking factor for the practical application of reusable static analyses.

More related, albeit focusing on extensibility rather then reusability, are the next two approaches. First, *Decorated Attribute Grammars* proposed by Kats, Sloane, and Visser employ language agnostic decorators upon domain-specific ASTs to simplify the specification of standard analysis tasks. Arguably, decorators can be used to specify a mapping, yet our approach can deal efficiently with arbitrary problem-specific data structures. Second, Söderberg, Ekman, Hedin, and Magnusson presents a JastAdd-based approach to perform simple control flow, dataflow and dead assignment analyses on Java programs also employing the *ExtendJ* extensible Java compiler. While these analyses operate on the Java AST, they could be extracted into reusable language-independent aspects following our approach. Notably though, extracting the control flow analysis would require a considerable effort.

Aside from all that, there a several approaches and tools for static analysis of specific target languages [18]. From these approaches, some support adding advanced static analysis. For instance, Bodden included inter-procedural flow analysis into *Soot* for Java programs [7]. This analysis generates an IDFS problem, which can be efficiently solved by a custom solver. While the solver could be generic, most of the implementation effort is hidden in the domain-specific IDFS generation. Another example, is the *StaRVOOrS* tool [8] for combined static and runtime verification of Java. It relies on a specification language for static and runtime verification of data flow and control flow analysis [2]. Similar to Soot, they generate a runtime specification with pre/post-conditions for the deductive theorem prover *KeY* and the runtime verification tool *Larva*. These approaches present a multi-staged transformation, which could also be facilitate using our approach, generating the required formal specifications.

Last but not least, two other approaches for the analysis of programming languages are distantly related: *SOUL* [16] and *Cubix* [25]. In *SOUL*, Smalltalk and Java programs are transformed into logic predicates, which are later used to, e.g., detect use or absence of design patterns. *Cubix* is a framework to specify and run language-independent transformations while still keeping language-specific features to retain as much as possible of the original source. Both approaches maintain certain links between the language-independent and the language-specific parts of the supported base languages, and both focus on programming languages rather than on DSLs.

In contrast, our approach also allows for reusing analyses for DSLs, however, is not focussed on transformation but mainly on analysis.

## 9 Conclusion

In this paper, we discuss why and how to construct reusable static analysis. The presented work illustrates how higher-order attributes can be utilized to map domain-specific concepts of a DSL to domain-independent data structures of static analysis algorithms. These algorithms, in turn, only operate on their data structures, and can thus be reused between different DSLs. Although we concede that this approach is not new, as it was pioneered by Joao Saraiva [36], our work shows both its practicality and feasibility by utilizing the state-of-the-art *Reference Attribute Grammar* system *JastAdd* [21] and its *Relational* extension [32]. In detail, we apply the approach to three different languages, a very simple state machine language, the very complex modelling language Modelica, and the general purpose language Java using two different analyses for reuse both across and within languages. Additionally, we systematically evaluated the performance of our approach compared to a reference implementation by analysing the dependencies of each Java project in the *QualitasCorpus*. Overall, our approach reduces the effort of adding and reusing static analysis between different DSLs, while only incurring a negligible performance overhead.

We acknowledge that not every type of static analysis can be performed efficiently and elegantly using RAGs. Therefore, we not only want to examine the applicability for other kinds of languages and analyses, but also plan to investigate how the approach can be used to efficiently construct input formats for specialized external tools. Additionally, we want to explore if incremental evaluation of RAG attributes can be utilized to allow incrementally rewrites of the domain-specific data structure whilst automatically recomputing the analysis.

**Acknowledgements**   This work is partly supported by the German Research Foundation (DFG) in the project "RISCOS" and as part of Germany's Excellence Strategy – EXC 2050/1 (CeTI), and by the German Federal Ministry of Education and Research within the projects "OpenLicht" and "KASTEL".

## A  Listings

### A.1  Cycle Detection

■ **Listing 13**  Cycle detection attributes defined for a dependency graph

```
1  syn Set<Set<Component>> DependencyGraph.SCC() {
2    Map<Component, Set>  visited = new HashMap<>();
3    LinkedList<Component> locked = new LinkedList<>();
4    for (Component c : getComponentList())
5      if (!visited.containsKey(c))
6        c.visit(visited, locked);              // forward search
7    for (Component c : locked)
8      if (visited.get(c) == null)
9        c.assign(visited, new HashSet());      // backward search
10   return new HashSet(visited.values());
11 }
12 void Component.visit(Map<Component, Set> visited, LinkedList locked) {
13   visited.put(this, null);
14   for (Component c : getFromList())
15     if (!visited.containsKey(c))
16       c.visit(visited, locked);
17   locked.addFirst(this);
18 }
19 void Component.assign(Map<Component, Set> visited, Set scc) {
20   scc.add(this);
21   visited.put(this, scc);
22   for (Component c : getToList())
23     if (visited.get(c) == null)
24       c.assign(visited, scc);
25 }
```

■ **Listing 14**  The relational NTA to compute the Java type dependency graph

```
1  syn lazy DependencyGraph Program.typeDependencyGraph() {
2    DependencyGraph dg = new DependencyGraph();
3    dg.setProgram(this);
4    Map<TypeDecl,Component> componentMap = new HashMap<>();
5    for (TypeDecl d: typeDecls()) {
6      TypeComponent c = new TypeComponent();
7      c.setTypeDecl(d);
8      componentMap.put(d, c);
9      dg.addComponent(c);
10   }
11   for (Component c: dg.getComponentList())
12     for (TypeDecl d: c.asTypeComponent().getTypeDecl().typeUses())
13       c.addTo(componentMap.get(d));
14   return dg;
15 }
16 coll HashSet<TypeDecl> Program.typeDecls() root Program; // collect all
17 TypeDecl contributes this to Program.typeDecls();        // type decls
18 coll HashSet<TypeDecl> TypeDecl.typeUses() root TypeDecl;
19 TypeAccess contributes decl()                      // collect all type uses
20   when program().typeDecls().contains(decl()) // of collected decls
21   to TypeDecl.typeUses() for hostType();       // in containing type decl
```

## A.2 Shadowing Analysis

This section contains a slightly simplified variant of the scope analysis for *JModelica*, for the complete, executable code and examples, please refer to the repository.[6]

■ **Listing 15**  Grammar of a scope tree data structure

```
1  RootScope : Scope;
2  abstract Element;
3  Declaration:Element ::= <Name:String>;
4  Scope:Element ::= Element*;
5  rel Scope.inheritedScope* -> Scope;
```

■ **Listing 16**  Shadowing analysis defined on the scope tree grammar

```
1  aspect Shadowing {
2    coll HashSet<ShadowFinding> RootScope.variableShadowings() root RootScope;
3    Declaration contributes new ShadowFinding(shadowed(), this)
4        when isShadowing()
5        to RootScope.variableShadowings();
6
7    syn Declaration Declaration.shadowed() = shadowedBy(asDeclaration());
8
9    inh Declaration Element.shadowedBy(Declaration shadower);
10   eq Scope.getElement().shadowedBy(Declaration shadower) =
11       shadowedLocally(shadower);
12
13   syn Declaration Scope.shadowedLocally(Declaration shadower) {
14     // first look in the current scope
15     for (Declaration decl : declarations())
16       if (decl != shadower && decl.getName().equals(shadower.getName()))
17         return decl;
18
19     // then look in the inherited scopes
20     for (Scope inherited : getInheritedScopeList()) {
21       Declaration shadowed = inherited.shadowedLocally(shadower);
22       if (shadowed != null) return shadowed;
23     }
24     return (isRootScope()) ? null : shadowedBy(shadower);
25   }
26
27   syn boolean Declaration.isShadowing() = shadowed() != null;
28
29   // Another analysis for repeated declarations
30   // within the same scope could be added here.
31 }
```

■ **Listing 17**  Mapping relations grammar from Modelica to scope trees

```
1  rel ScopeTree.SourceRoot -> SourceRoot;
2  ClassDeclScope : Scope;
3  rel ClassDeclScope.classDecl -> SrcClassDecl;
4  ComponentDeclaration : Declaration;
5  rel ComponentDeclaration.componentDecl -> SrcComponentDecl;
```

---

[6] https://git-st.inf.tu-dresden.de/jastadd/reusable-analysis

## Reusing Static Analysis across Different DSLs using RAGs

■ **Listing 18**   Transformation attributes from Modelica to a scope tree

```
1  aspect ModelicaToScopeMapping {
2    /** a relational nta attribute to compute the scope tree */
3    syn lazy RootScope SourceRoot.scopeTree() {
4      RootScope tree = (RootScope) scope();
5      for (SrcClassDecl decl : topLevelClasses()) // add top-level classes
6        tree.addElement(decl.scope()); // using the collection declared in l.14
7      tree.updateInheritance(); // traverse scopes and add inheritance relations
8      return tree;
9    }
10
11   /** a relational nta collection attribute to compute scopes */
12   coll Scope ASTNode.scope() [asScope()] with addElement root SourceRoot;
13   SrcClassDecl contributes scope()
14       when isInnerClass()
15       to ASTNode.scope()
16       for containingScope();
17   SrcComponentDecl contributes asDeclaration()
18       to ASTNode.scope()
19       for containingScope();
20
21   /** helper methods to add inheritance relations */
22   public void Scope.updateInheritance() {
23     for (Element element : getElementList())
24       if (element.isScope()) element.asScope().updateInheritance();
25   }
26   public void ClassDeclScope.updateInheritance() {
27     for (SrcExtendsClause extendsClause : getClassDecl().superClasses()) {
28       SrcClassDecl superClass = extendsClause.getSuper().findClassDecl();
29       if (superClass != null) addInheritedScope(superClass.asScope());
30     }
31     super.updateInheritance();
32   }
33 }
```

■ **Listing 19**   Constructor NTAs for a Modelica scope tree

```
1  aspect MappingConstructors {
2    syn lazy RootScope SourceRoot.asScope() {
3      RootScope tree = new RootScope();
4      tree.setSourceRoot(this);
5      return tree;
6    }
7    syn lazy ClassDeclScope SrcClassDecl.asScope() {
8      ClassDeclScope scope = new ClassDeclScope();
9      scope.setClassDecl(this);
10     return scope;
11   }
12   syn lazy ComponentDeclaration SrcComponentDecl.asDeclaration() {
13     ComponentDeclaration decl = new ComponentDeclaration(getName().getID());
14     decl.setComponentDecl(this);
15     return decl;
16   }
17 }
```

■ **Listing 20**  Transformation helper attributes

```
1  aspect ScopeGenerationAttributes {
2    /** inherited attributes to determine the scope an AST element */
3    inh lazy ASTNode SrcClassDecl.containingScope();
4    eq SrcClassDecl.getChild().containingScope() = this;
5    inh lazy ASTNode SrcComponentDecl.containingScope();
6    eq SrcForStmt.getChild().containingScope() = this;
7
8    // other navigation attributes omitted or in JModelica base implementation
9  }
```

## B  Evaluation Tables

■ **Table 1**  Package analysis of the eight largest Java projects in Qualitas Corpus [41]

| Scenario | Files | Graph Size | Direct Median (ms) | Reusable Median (ms) | Overhead (%) |
|----------|-------|------------|--------------------|--------------------|-------------|
| Lucene | 3036 | 1555 | 3195 | 3167 | −0.88 |
| Azureus | 3319 | 4347 | 3464 | 3437 | −0.79 |
| Spring Framework | 4202 | 2484 | 3121 | 3171 | 1.58 |
| Hibernate | 6230 | 4607 | 3604 | 3519 | −2.42 |
| JBoss | 6809 | 2868 | 4077 | 4013 | −1.60 |
| GeoTools | 7134 | 3815 | 5952 | 6033 | 1.34 |
| Eclipse | 22 634 | 22 784 | 29 333 | 28 456 | −3.08 |
| NetBeans | 32 647 | 38 330 | 51 968 | 51 590 | −0.73 |

■ **Table 2**  Type analysis of the eight largest Java projects in Qualitas Corpus [41]

| Scenario | Files | Graph Size | Direct Median (ms) | Reusable Median (ms) | Overhead (%) |
|----------|-------|------------|--------------------|--------------------|-------------|
| Lucene | 3036 | 37 683 | 4818 | 4689 | −2.75 |
| Azureus | 3319 | 41 084 | 5613 | 5465 | −2.71 |
| Spring Framework | 4202 | 31 938 | 3777 | 3684 | −2.52 |
| Hibernate | 6230 | 41 658 | 5153 | 4910 | −4.95 |
| JBoss | 6809 | 26 536 | 4816 | 4736 | −1.69 |
| GeoTools | 7134 | 55 460 | 8474 | 8323 | −1.81 |
| Eclipse | 22 634 | 313 879 | 49 150 | 48 452 | −1.44 |
| NetBeans | 32 647 | 395 999 | 79 506 | 78 707 | −1.02 |

## About the authors

**Johannes Mey** is a research assistant and PhD student at the chair of software technology at Technische Universität Dresden. His research focuses on reference attribute grammars an their application in various fields. Besides static program analysis, these include model transformations and adaptive software systems. Furthermore, he investigates the relation between reference attribute grammars and conceptual models. Contact him at johannes.mey@tu-dresden.de.

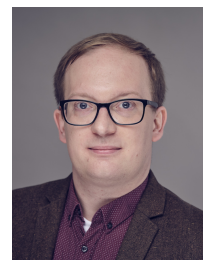

**Thomas Kühn** is a post-doc at the Software Design and Quality Group at Karlsruhe Institute of Technology. His research focuses one new ways to model and program future software systems challenged by increased complexity, heterogeneity, rate of change and longevity. As a result, he developed a family of role-based modelling and a family of role-oriented programming languages supported by a feature-aware modelling editor and a basic IDE, respectively. Currently, he improves tool support for view-based, model-driven software development building on the Vitruvius approach. Contact him at thomas.kuehn@kit.edu.

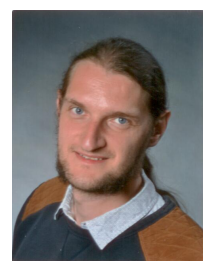

**René Schöne** is a research assistant and PhD student at the chair of software technology at Technische Universität Dresden. His research and PhD focuses on the application of reference attribute grammars for models@run.time currently within the domain of smart home. Challenges there include adequate modelling of domains, abstraction of and connection to real hardware devices, and efficient analyses in the present of frequent model updates. Contact him at rene.schoene@tu-dresden.de.

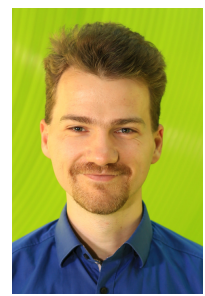

**Uwe Aßmann** holds the Chair of software technology at the Technische Universität Dresden. He has obtained a PhD in compiler optimization and a habilitation on *invasive software composition* (ISC), a composition technology for code fragments enabling flexible software reuse. ISC unifies generic, connector-, view-, and aspect-based programming for arbitrary program or modelling languages. Since 2013, he is deputy of the DFG Research Training Group *Role-oriented Software Infrastructures* (RoSI), which develops new techniques for context-adaptive software, from language and application design to run time (rosi-project.org). Contact him at uwe.assmann@tu-dresden.de.

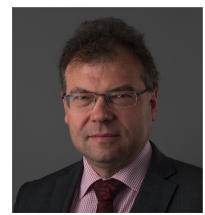